# Electronic structure of (Ga,Mn)As revisited

J. Kanski[a], L. Ilver[a], K. Karlsson[b], I. Ulfat[c], M. Leandersson[d], J. Sadowski[d,e], I. Di Marco[f]

[a]Department of Applied Physics, Chalmers University of Technology, SE-412 96 Göteborg, Sweden, [b]Department of Engineering Sciences, University of Skövde, SE-541 28 Skövde, Sweden, [c]Department of Physics, University of Karachi, Karachi 75270, Pakistan, [d]MAX IV Laboratory, Lund University, SE-221 00 Lund, Sweden, [e]Institute of Physics, Polish Academy of Sciences, al. Lotników 32/46, PL-02-668 Warszawa, Poland, [f]Department of Physics and Astronomy, Uppsala University, Box 516, SE-75120, Uppsala, Sweden

Corresponding author:
Janusz Kanski, Dept. of Physics, Chalmers University of Technology, SE 41296 Gothenburg, Sweden. e-mail: janusz.kanski@chalmers.se, tel. +46 317723313

Abstract

The detailed nature of electronic states mediating ferromagnetic coupling in dilute magnetic semiconductors, specifically (Ga,Mn)As, has been an issue of long debate. Two confronting models have been discussed emphasizing host band vs. impurity band carriers. Using angle resolved photoemission we are for the first time able to identify a highly dispersive Mn-induced energy band in (Ga,Mn)As. Our results show that the electronic structure of the (Ga,Mn)As system is significantly modified from that of GaAs throughout the valence band. Close to the Fermi energy, the presence of Mn induces a strong mixing of the bulk bands of GaAs, which results in the appearance of a highly dispersive band in the gap region of GaAs. For Mn concentrations above 1% the band reaches the Fermi level, and can thus host the delocalized holes needed for ferromagnetic coupling. Overall, our data provide a firm evidence of delocalized carriers belonging to the modified host valence band.

Introduction

Although more than 20 years have passed since the first synthetization of a III-V-based dilute magnetic semiconductor (1), implementation of these materials in everyday spin-based electronics is as elusive as ever because the ferromagnetic transition temperature is much lower than desired. Rather remarkably, the physical origin of the ferromagnetic state is still debated, even for the prototype dilute magnetic semiconductor (Ga,Mn)As. A wealth of experimental data suggests that the magnetic coupling is mediated by spin-polarized holes, but the actual character of these holes has become an issue of fierce debate. Two main scenarios are discussed: acceptor induced holes in the host valence band versus holes in a more or less detached impurity band. Experimental evidence for the existence of an impurity band based on optical properties has been presented (2), though later studies suggested that these data are also consistent with the host valence band model (3). Support for an impurity band scenario is also obtained from resonant tunneling experiments on quantum well structures (4) and from channeling in combination with magnetization, transport, and magneto-optical experiments (5). In this last work the location of the Fermi level within the impurity band is emphasized to play a crucial role in determining the Curie temperature ($T_C$). Studies based on photoemission, instead, point to the coexistence of coupling mechanisms in the impurity band and host valence band models (6, 7). The role of delocalized Mn-derived states near the top of the valence band has been emphasized in photoemission studies using very high photon energies (8), and the presence of delocalized Mn d-states has been inferred from observation of screening effects in core level spectra correlated with magnetic properties (9). As will be shown here the direct contribution to the electronic structure of the Mn 3d states is not as significant as suggested in these studies. Instead the role of the Mn impurities is to perturb the host valence bands. In a very recent photoemission study (10) it was found that the electronic structure of (Ga,Mn)As is heavily perturbed by disorder in the region of valence band maximum (VBM) relative to that of GaAs. However, these experiments were carried out only with radiation from a He discharge lamp and did not capture any of the main observations of the present angle resolved photoemission study, which instead uses synchrotron radiation.

Experimental aspects

While photoemission is the most direct probe of electronic states, its applicability is hampered by its intrinsic surface sensitivity: well-defined, atomically clean samples are required. This is not an issue in situations where the surface can be prepared by e.g. ion etching and annealing, but in the present case such treatment is prohibited because (Ga,Mn)As undergoes phase separation at temperatures above 300 °C. Indeed, in an earlier study (11) it was demonstrated that the electronic structure is modified by annealing, the most obvious effect being a shift of the Mn 3d binding energy from around 3.2 eV (for as-grown material) (7) to 4.3 eV (after post-growth treatment) (12). Interestingly (and surprisingly), the latter value is still quoted in literature (see e.g. 13), which further adds to a confusing discussion. Even if phase separation can be avoided by annealing at lower temperature (14), etching of a ternary system like (Ga,Mn)As may modify the surface composition and morphology in an uncontrolled way. An alternative is to use As capping to protect the surface against contamination during transfer between the growth and analysis units (15). This again is a very delicate method, since the capping must be sufficiently thick to serve its purpose (typically 400 nm, see ref. 16). The As capping applied in ref. 15 was only 0.5-1 nm thick, and the XAS reported in ref. 15 did indeed show the structures characteristic for an oxidized sample (17). On the other hand, a sufficiently thick As capping would have to be removed by heating, in which case an additional complication is unavoidable: during post-growth annealing interstitial Mn will diffuse to the surface and react with As to form MnAs overlayer/particles (18). A different approach to avoid problems with the surface is to reduce the surface sensitivity of the photoemission spectroscopy by using a sufficiently high photon energy (6). As will be shown in the present study, however, features of major interest are confined near the centre of the Brillouin zone, and would therefore appear in an angular range of less than 0.5° from the surface normal at the photon energy used in ref. 6. Clearly, these features could not be resolved in the experimental conditions applied in ref. 6.

The only safe way to avoid complications with surface preparation is transfer of samples between the growth and analysis systems in ultrahigh vacuum (UHV). This is the strategy adopted in the present work, and was also followed in another very recent study (10). The fact that the results presented here have not been found in any previous study shows unequivocally that sample handling is a decisive issue not only for magnetism (19) but also for the details of the electronic structure. The present data allow us to examine the nature of the electronic states close to the Fermi level and for the first time identify a highly dispersive Mn-induced band which can provide the delocalized holes needed for the ferromagnetic coupling.

Results and discussion

Two sets of experimental data are presented here, one obtained at MAX IV laboratory beamline I3, where a photoelectron spectrometer is connected to an MBE system, the other at the Swiss Light Source (SLS) ADRESS beamline. In the latter case the samples were transported in a UHV suitcase from the MBE system at MAX-lab. The (Ga,Mn)As layers were grown on n-type GaAs(100) substrates and the Mn concentration was determined using RHEED oscillations with an accuracy better than 0.1% (20). To allow detailed comparison of spectra from GaAs and (Ga,Mn)As a mask was used during the growth, leaving a part of the substrate with clean GaAs. In this way spectra from the two materials could be recorded under identical conditions. The SLS data discussed here were recorded with the sample at around 200 K, while the MAX data were obtained at room temperature. In both cases the temperatures were far above the Curie temperature of as-grown samples, which is typically below 50 K. All samples with Mn concentrations above 0.5% showed (1x2) LEED patterns, while for pure GaAs the LEED pattern was c(4x4) (Fig. S1).

Figs. 1a and 1b show photoemission intensity distributions obtained at SLS with circularly polarized 453 eV photons. On the whole, the data from GaAs and (Ga,Mn)As are similar except in the vicinity of VBM, where the emission from (Ga,Mn)As appears fuzzy. This region will be discussed using the MAX-lab data, which were recorded with much better angular resolution and better statistics. There are other, less obvious differences, that are disclosed via intensity profiles. In Fig. 1c a pair of profiles is displayed, selected such that the in-plane momentum separations ($\Delta k_{//}$) between the light hole (LH) branches is the same (at the dashed lines in Figs. 1a and 1b). We find that $\Delta k_{//}$ between corresponding spin-orbit (SO) branches is somewhat larger for (Ga,Mn)As than for GaAs. In Fig. 2 we show how this difference in $\Delta k_{//}$ develops along the bands in the binding energy range 1.5 – 4.0 eV. A corresponding plot of $\Delta k_{//}$ between heavy and light hole bands does not reveal any significant difference between the two materials (Fig. S3). Clearly, the bulk band structure of GaAs is modified by the introduction of Mn in a non-trivial way, i.e. more than a rigid shift due to p-doping. Regarding the origin of this modification, we note that early photoluminescence data from heavily Zn-doped GaAs showed a shift of the emission involving the SO band, that was ascribed to a smaller contribution to the spin-orbit energy in the dopant (21). In analogy, it can be expected that the SO splitting should be reduced by replacing Ga with Mn in (Ga,Mn)As.

It is noted in Fig. 1 that the LH and SO bands are excited with approximately the same probability over a range of binding energies. This can be understood as an effect of final state lifetime broadening. The intensity distribution then reflects the projected density of states. As illustrated schematically in the inset in Fig. 1c, the projection of a spherical momentum distribution (i.e. E~|**k**|$^2$) is a circle with maximum density of states at its periphery, and the intensity profile takes the experimentally observed "suspension bridge" shape.

We now turn to the MAX-lab results. As for the SLS data, the valence band region is characterized by overall similarities between (Ga,Mn)As and GaAs and the SO band of (Ga,Mn)As is shifted up in energy (Fig. S2). A deformation of the SO band was also reported in an early photoemission study of (Ga,Mn)As (14), though the shift was in the opposite direction to that found here. The cause for this discrepancy is not clear, but the energy alignment is an obvious issue of concern - as described below, we have chosen to align the X$_3$ critical points, while in ref. 14 (and likewise in ref. 6) the Fermi energy was used as a reference. The latter is obviously misleading because the doping situations in the two materials are very different. In Fig. 3 we show intensity distributions in the VBM region. As expected, the Fermi level in GaAs is pinned near midgap and the gap region is completely free from photoelectrons (Fig. 3b). For (Ga,Mn)As the emission extends towards higher energies, as seen in Fig. 3c. Due to the rapidly falling intensity, it is not possible to display the details without totally overexposing the image. To get around this complication we composed an image from slices, each arbitrarily adjusted with respect to threshold and saturation levels. The resulting image, displayed in Fig. 3d, clearly shows that the spectral tailing actually reflects a well-defined energy band that reaches the Fermi level. *Thus, unlike all previous studies, we are able to directly detect delocalized electron states at the Fermi level that are specific for (Ga,Mn)As*. In Fig. 3d we also indicated the VBM position in GaAs (dotted line), taking into account the different pinning situations. This energy was estimated using literature data (22, 23), according to which the separation between the $X_3$ point and VBM is in the range 6.7 – 6.9 eV. The dotted line marks the highest possible position based on these data, so it can be safely concluded that the narrow band extends into the band gap region of GaAs. It is motivated to emphasize that one reason why this band has eluded detection in previous studies (apart from the sample preparation issue discussed above) is its low intensity in combination with steep dispersion: in regular energy distribution curves the structure appears as a weak shoulder on a tailing background, and experiments with limited angular resolution the dispersive character will be hardly discernible. Instead a slightly increased intensity will manifest itself as a peak in difference spectra (6, 15).

Having established the existence of a dopant-induced energy band above VBM, we proceed to examine its properties. Of immediate concern is the possibility that it may reflect a surface state. Within the photon energy range 20 – 35 eV, where the band is observed, we found no significant dependence of momentum along the surface normal. While this is normally a reliable signature of a surface state, several other observations contradict such interpretation. First, the band is not confined to the band gap region, but can be followed well below VBM. Second, no asymmetry was observed that could be connected with the surface reconstruction (as is the case for e.g. the GaAs(100)-2x4 surface (24)). Third, a well-defined and rapidly dispersing surface state band would require a well-ordered surface with long-range coherence. However, several studies (Fig. S1, 25) have shown that the (1x2) reconstructed surface is characterised by disorder. Fourth, the band has been found quite stable against surface contamination (adsorption of residual CO and $N_2$) and is clearly observed even when the most prominent bulk derived features are strongly attenuated. All this leads us to conclude that the Mn-induced band is not a surface state but a feature of the bulk electronic structure. It is appropriate at this point to note that also in the above-mentioned angle-resolved photoemission study (14) a structure lacking dispersion along the surface normal was reported. In contrast to the Mn-induced band discussed here, however, that structure was located well below the Fermi level (0.5 – 1.0 eV), and did not show any in-plane dispersion. We tentatively ascribe the difference relative the present results to the ion etching treatment during surface preparation (14).

The above observations can be related to the magnetic properties of (Ga,Mn)As. It is known (26) that (Ga,Mn)As is ferromagnetic only at Mn concentrations higher than approximately 1% and that $T_C$ is proportional to the density of holes. Our 1.2% sample is a borderline case with a measured $T_C$ of around 10 K. In Fig. 4 we compare photoemission data for different Mn concentrations. For the 0.5% Mn sample the band does not reach the Fermi level, and no ferromagnetism was found. With 5% Mn the band appears broader, which can be understood as an upwards shift in energy. As a result the density of states at the Fermi level is increased, and indeed the recorded $T_C$ for this sample was around 50 K. The concentration dependence of the Mn-induced band is obviously matching the known magnetic properties, so the band reaching the Fermi level is an obvious candidate for hosting the delocalized holes needed for ferromagnetic coupling.

Features in the VBM region resembling those found here have been reported in a couple of theoretical studies (27, 28). In these works an excursion of a host-derived majority spin band is predicted above VBM. However, these studies address (Ga,Mn)As in its ferromagnetic state, while the data discussed here were recorded well above $T_C$. Using the ferromagnetic phase to understand the paramagnetic phase of (Ga,Mn)As is fully justified only if their electronic structures are qualitatively similar. We note that electronic structure calculations of the paramagnetic state in the disordered local moments (DLM) picture give substantially the same magnetic local moment and (spin-integrated) spectral properties of the ferromagnetic phase (29). Although the DLM picture involves only an approximate treatment of spin-fluctuations, the absence of drastic changes in the electronic structure across the ordering temperature is not uncommon for systems where the magnetic moments arise from strongly localized electrons (30). The localized nature of the Mn-3d states in (Ga,Mn)As is indeed widely accepted (18), and is also suggested by the multiplet-like spectrum reported recently (7).

We return now to the question concerning the excitation of the Mn-induced band. In "regular" crystal momentum assisted photoemission, dispersive bulk states are observed at a fixed in-plane momentum and appear at different binding energies in spectra excited with different photon energies. Reversibly, the lack of such photon energy dependence is a typical property of a surface state. As already discussed, various observations contradict a surface state interpretation in the present case. A striking observation is the very low spectral intensity, about two orders of magnitude smaller than that of the surface state of the GaAs surface. Since we associate the band with Mn impurities, it is natural to suspect that the low intensity may be directly related to the low density of impurity atoms. However, apart from the fact that we do not observe any clear proportionality between the intensity and Mn concentration over a range of 1-5 %, this would not explain the lack of dispersion along the surface normal. Alternatively, the low intensity can be taken as an indication of a basically different excitation mechanism from that in regular photoemission. A mechanism that is usually ignored in the analysis of photoemission spectra is the one based on the change of the photon field in the surface region, so called surface photoemission. Surface photoemission has been discussed extensively in the past (31), mainly in connection with excitation of sp-bands in metals. By this mechanism the momentum selectivity along surface normal is relaxed while the in-plane momentum is preserved. The emission should then reflect the projected density of states, as discussed in connection with the SLS data, and the intensity distribution should appear as a dispersing band just as the SO band in Fig. 1. If this is correct, corresponding emission should also be found for GaAs. To test this hypothesis we report intensity distributions from the VBM region of pure GaAs (Fig. 5a) and (Ga,Mn)As with only 0.4% Mn (Fig. 5b). The data are displayed in the 2$^{nd}$ derivative mode with high intensity represented by bright colour. For GaAs we have indicated the bulk bands using effective masses and SO splitting from literature (32). A feature of particular interest in Fig. 5a is the bright spot at normal emission around 1.3 eV binding energy. This spot marks the top of a triangular field, which can be followed about 1 eV down in energy. As suggested by the dashed lines, we associate the bright spot with the top of the SO band. One can also discern a bright path coinciding with the downward dispersing SO band. The shaded triangular field, reflecting a relatively high intensity, is contained within the SO band in much the same way as the field between the SO branches in Fig. 1. We can conclude that the intensity distribution can be explained as projected density of states, supporting the above interpretation of the excitation mechanism. For (Ga,Mn)As the intensity distribution contains a similar triangular bright region, Fig. 5b, but the spot marking the top of the triangle is missing and the triangular region appears to extend through the VBM into the band gap of GaAs. The data indicate that the band structure of (Ga,Mn)As cannot be considered as p-doped GaAs, but rather as a system in which the energy bands of the host material are intermixed and modified by the Mn impurities. A more detailed description of the band structure of (Ga,Mn)As requires a theoretical analysis that takes into account the impurities as well as the interaction with the hole gas.

While the connection between the magnetic properties and the dispersive Mn-induced band is indeed suggestive, the actual origin of the Mn-induced modifications remains to be clarified. From a theoretical point of view, in addition to the doping induced shift, changes in the electronic structure are due to the hybridization with impurity states and disorder. Moreover, an effect that is generally overlooked in the literature on dilute magnetic semiconductors is the interaction between host electrons and the hole gas. A well-known consequence of such interaction is a dopant-induced bandgap narrowing (33, 34, 35). Recalling that $Mn_{Ga}$ is an acceptor in GaAs, the hole density in samples with Mn concentration in the range of 1% is above $10^{20}/cm^3$ (even taking into account compensation due to Mn interstitials). With such strong doping the band gap narrowing is expected to be in the region of 100 meV. It is conceivable, therefore, that the present observations are, at least partly, due to the effects of the dopant-induced hole gas.

Summary

In conclusion, the present study provides two new observations with significant impact on the view on the electronic states in (Ga,Mn)As: a) we show that the band structure of the host material is modified in a non-trivial way, such that the bulk bands are modified over a wide energy range. In ref. 8 the largest Mn derived modification of the GaAs band structure was actually predicted in the binding energy region 2-4 eV, though this was not verified experimentally; b) most importantly, a highly dispersive Mn-induced energy band is found *above* the valence band maximum of the host material. The development of this band can be observed at Mn concentrations below 0.5%. For concentrations above 1% this band reaches the Fermi level (that is located in the band gap of GaAs) and can host holes mediating the ferromagnetism. This is the first time that such features are observed *directly* - previous photoemission experiments lacked the necessary angular resolution. Apart from these novel features of crucial importance, our data are in good agreement with the most recent photoemission measurement what regards the gross features, e.g. the binding energy of the main Mn 3d peak at around 3 eV (10, 15). However, while we find that the Fermi level is in the gap region of GaAs, in ref. 10 it is concluded to be deep below VBM in the LH/HH band. In comparison with other recent studies, in which modifications of the host valence band have been inferred (e.g. ref. 8), it is important to stress that the present data provide a qualitatively different picture: the modifications are not described as Mn-*derived* but Mn-*induced*. The distinction might appear subtle, but it is indeed significant. In the former case the modification is due to intermixing of Mn 3d states with host valence states, the Mn states retaining their Gaussian line shape, in the latter the Mn impurities induce changes in the host band structure. The present finding is of crucial importance, since it reconciles the successes obtained by the p-d Vonsovsky-Zener model of magnetism (19) with spectroscopic data favouring the valence band model. Our study also reveals that the host valence band is modified by Mn-impurities such that the effect of the dopants is not just a shift of the chemical potential. Furthermore, no evidence of a detached impurity band is found even for concentrations below 0.5%, which suggests that the host valence band model (delocalized holes) stays more or less valid till the Anderson metal-to-insulator transition.

The "Battle of the bands" (36) appears to be more complex than previously imagined. It is indeed surprising that after nearly two decades of studies by several groups the general picture of the situation can be changed so radically.

Acknowledgements

Valuable discussions with Olle Eriksson, Barbara Brena and Olle Gunnarsson are gratefully acknowledged. We also wish to thank Vladimir Strocov and Federico Bisti at SLS for important experimental assistance. J.K. and L.I. wish to express gratitude to Peter Apell for all support. JS acknowledges partial support from the research project No: 2014/13/B/ST3/04489 financed through the National Science Centre (Poland).

Figure Legends

Figure 1   Photoemission intensity distributions from a) GaAs(100)-c(4x4) and b) (Ga,Mn)As(100)-(1x2) excited with circularly polarized 453 eV photons. c) Intensity profiles extracted at the energies marked by dotted lines in a) and b). The inset shows a schematic model of the projected density of states (DOS) for a free-electron like band.

Figure 2   The $k_{//}$ separation between the two branches of the SO band as a function of the corresponding separation between the LH branches. The inserted binding energy scale refers to the GaAs data. The lines show parabolic fits to the respective data.

Figure 3   Photoemission intensity distributions the VBM region of a) GaAs and b) (Ga,Mn)As, excited with p-polarized 21 eV photons. c) The same data as a) but with a reduced threshold level. d) The same (Ga,Mn)As data as in b) but composed of slices with gradually reduced threshold level. The dashed line indicates the Fermi energy and the dotted line in d) represents the valence band maximum of GaAs as described in the text.

Figure 4   VBM data at room temperature from (Ga,Mn)As samples with a) 0.5% (excited with 24 eV photons), b) 1.2% and c) 5% Mn (both excited with 25 eV photons). The dashed line indicates the Fermi level. The Curie temperature for the sample with 1.2% Mn was around 10 K and around 55 K for the 5.5% sample. No ferromagnetism was found for the sample with 0.5% Mn.

Figure 5   VBM data at room temperature from a) GaAs and b) (Ga,Mn)As with 0.4% Mn, excited with 25 eV photon energy. The two figures are aligned at the $X_3$ density of states peaks. Intensity profiles are indicated in both cases at 1.5 eV binding energy. For GaAs we have indicated the bulk bands using effective masses and SO splitting from ref. 32.

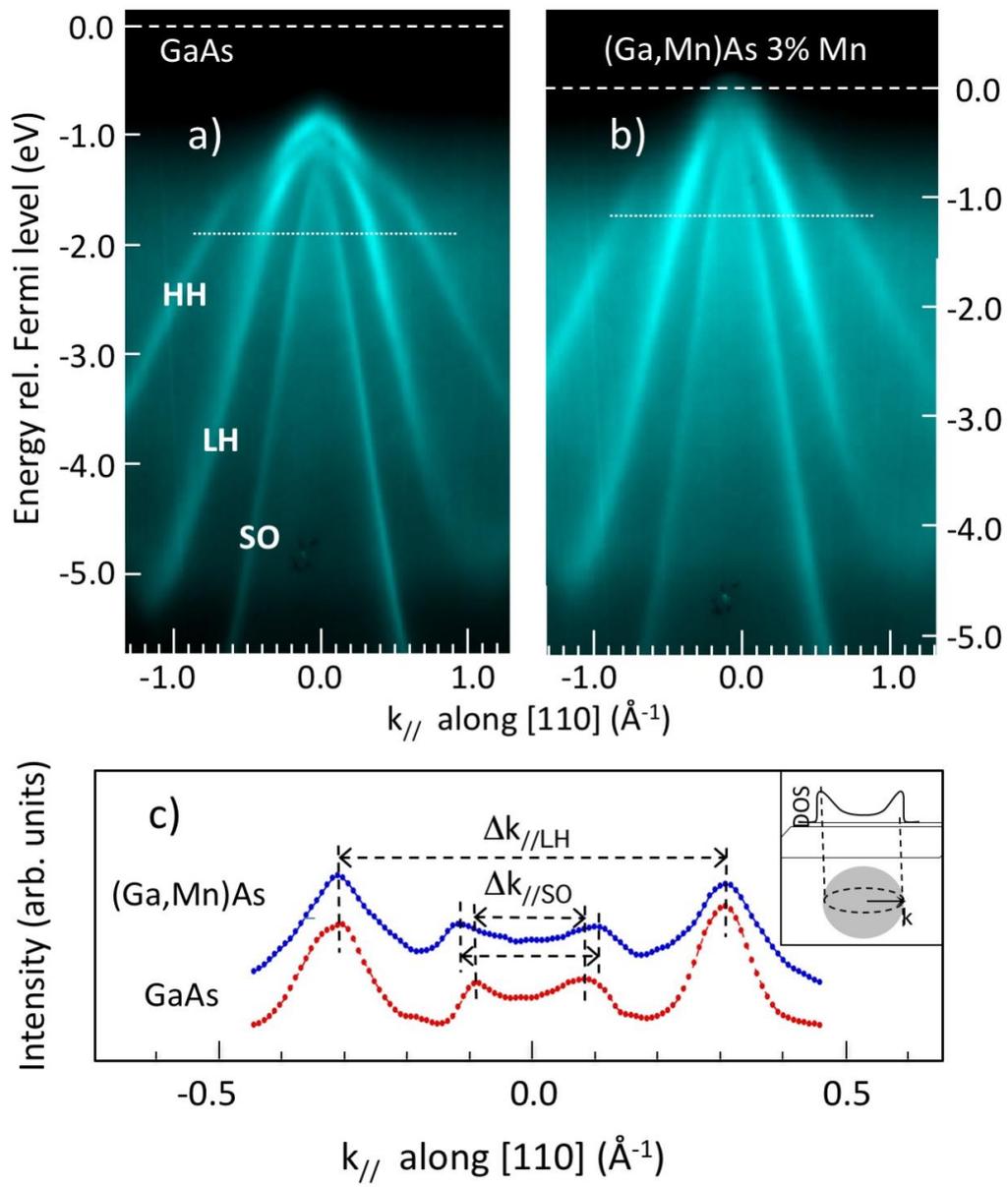

**Figure 1**

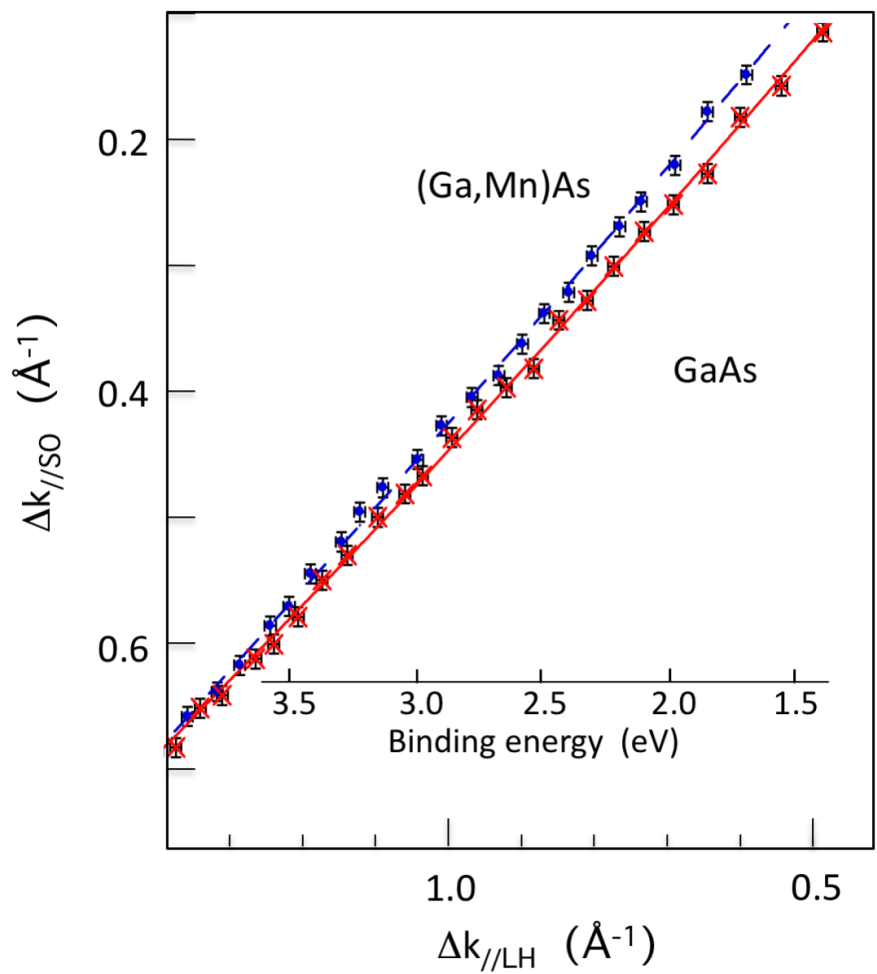

**Figure 2**

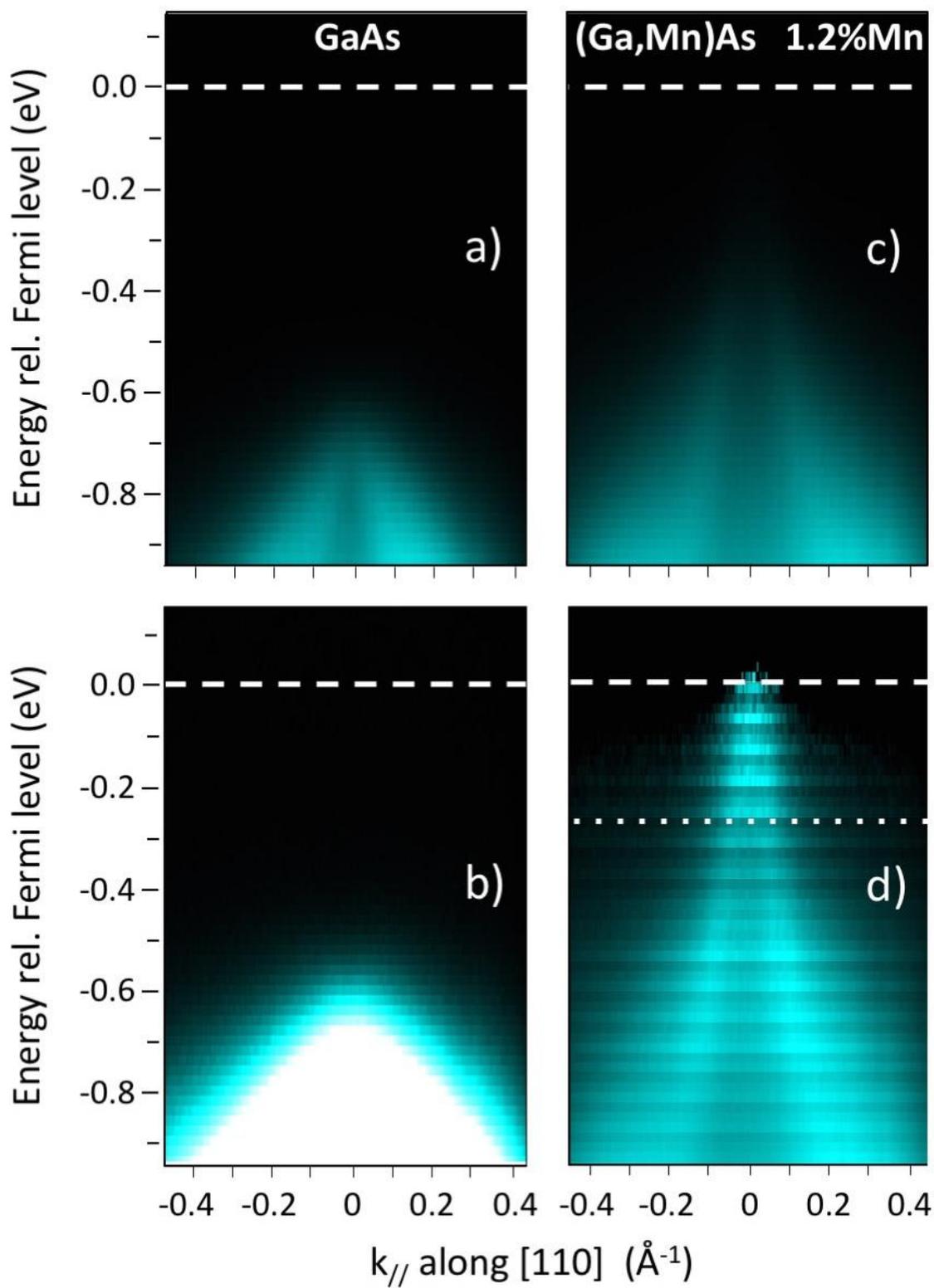

**Figure 3**

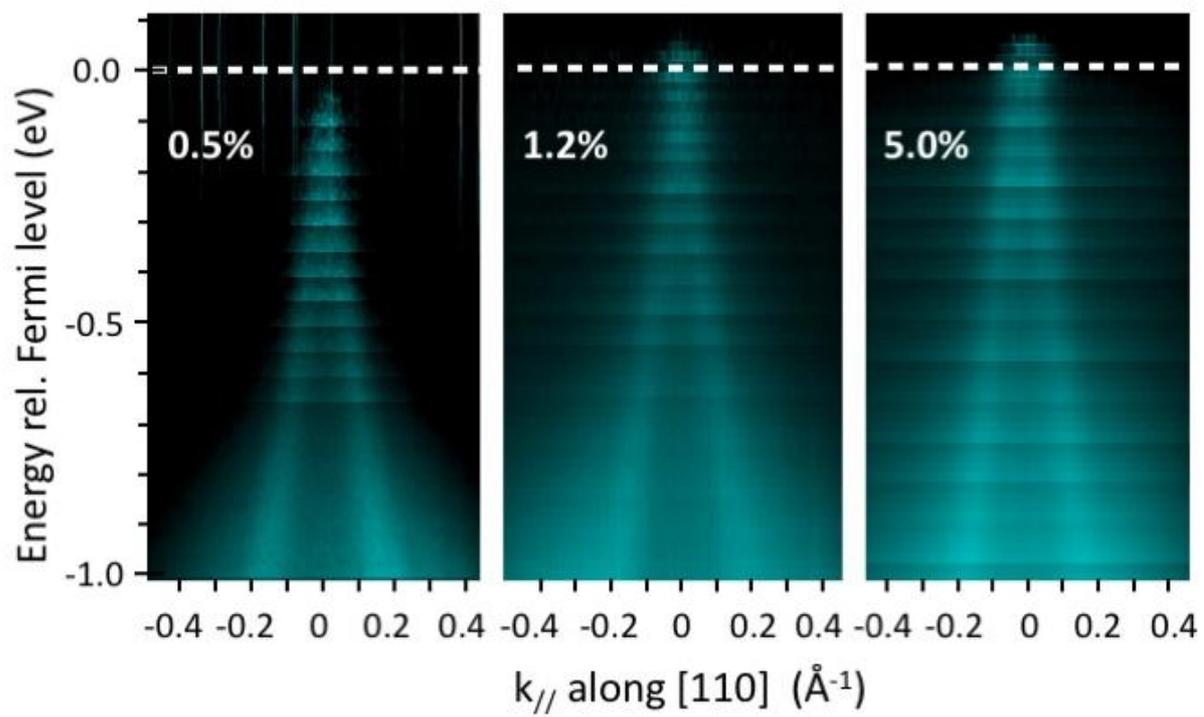

**Figure 4**

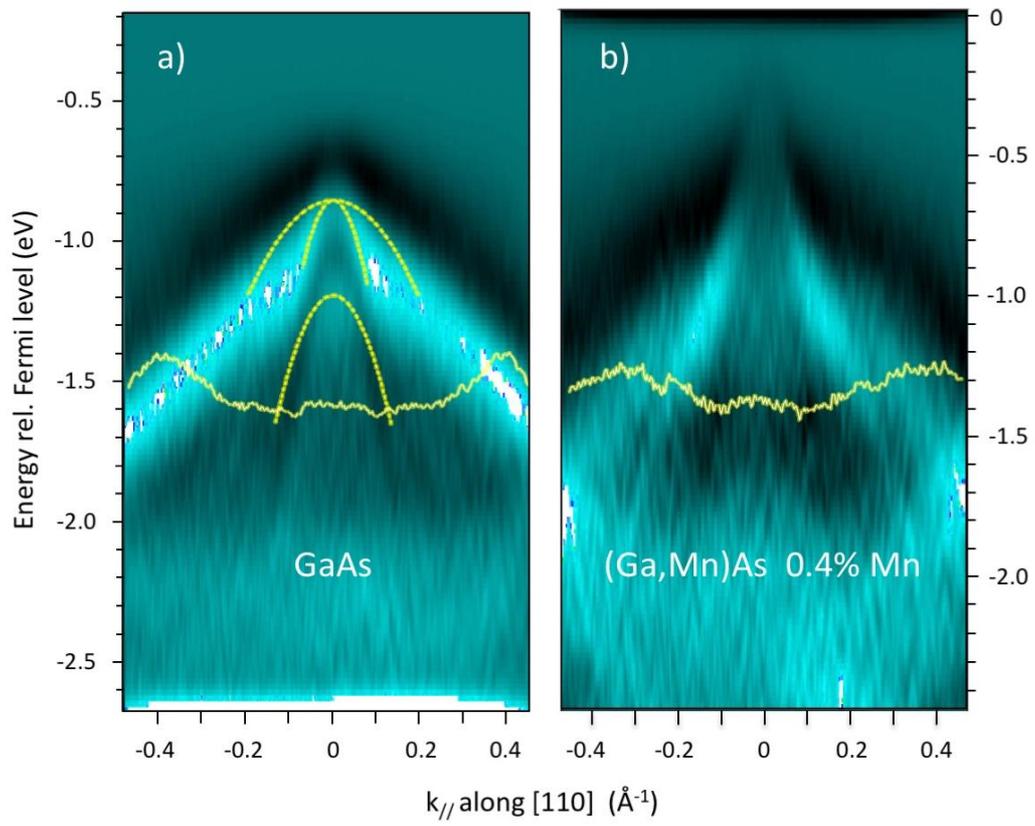

**Figure 5**

Supplemental information

During MBE growth of (Ga,Mn)As half-order streaks are observed in RHEED, indicating reconstruction along the <110> direction. The reconstruction is confirmed by LEED after transfer to the electron spectrometer system. However, as shown in Fig. S1, the LEED patterns of (Ga,Mn)As, obtained under identical conditions as those from GaAs, are characterized by a higher background, indicative of a relatively less ordered surface. At low kinetic energies some streaking is also observed (Fig. S1c), revealing preferential disorder along the <100> azimuth.

All photoemission spectra were excited with p-polarized light, incident at 12° and 17° grazing angles at MAX-lab and at SLS, respectively. The data discussed here were obtained with photon energies in the range 20-25 eV (at MAX-lab) and 450 eV (at SLS), which means approximately 1 nm electron mean free paths in both cases. For reliable detailed comparison of spectra from GaAs and (Ga,Mn)As the two samples were made on the same piece of GaAs wafer: after deposition of a GaAs buffer on the whole substrate, a mask was introduced to protect part of the surface during the subsequent deposition of (Ga,Mn)As. In this way spectra could be recorded from the two materials under identical conditions just by shifting the substrate 2-3 mm in front of the analyzer. In the comparison of spectra from GaAs and (Ga.Mn)As the $X_3$ density of states peak was used for energy alignment, as shown in Fig. S2.

One of our main findings is the modification of the host material bulk bands by the introduction of Mn. Specifically, we found a systematic change of the SO band relative the LH band (Fig. 2). As shown in Fig. S3, a corresponding plot of the HH band vs. LH band does not reveal any systematic change.

Figure Legends (SI)

Figure S1  LEED patterns from GaAs(001)-c( 4x4)  (a and b) and (Ga,Mn)As(001)-(1x2) (c and d). The electron energies are indicated above the pictures.

Figure S2  Second derivative presentations of valence band intensity distrubutions from a) GaAs(100)-c(4x4) and b) (Ga,Mn)As(100)-(1x2) with 1.2 % Mn. The spectra were excited with p-polarized 21 eV photons and recorded along the [-110] azimuth. The two spectra shown on the right side, were obtained by integrating the intensity distributions over ± 2° around the surface normal. Note that the width of the $X_3$ peak is not affected by introduction of Mn, but that the $\Delta$ band emission is shifted towards higher energy.

Figure S3  In-plane momentum separations between the branches of HH vs. LH bands in GaAs (circles) and (Ga,Mn)As (crosses).

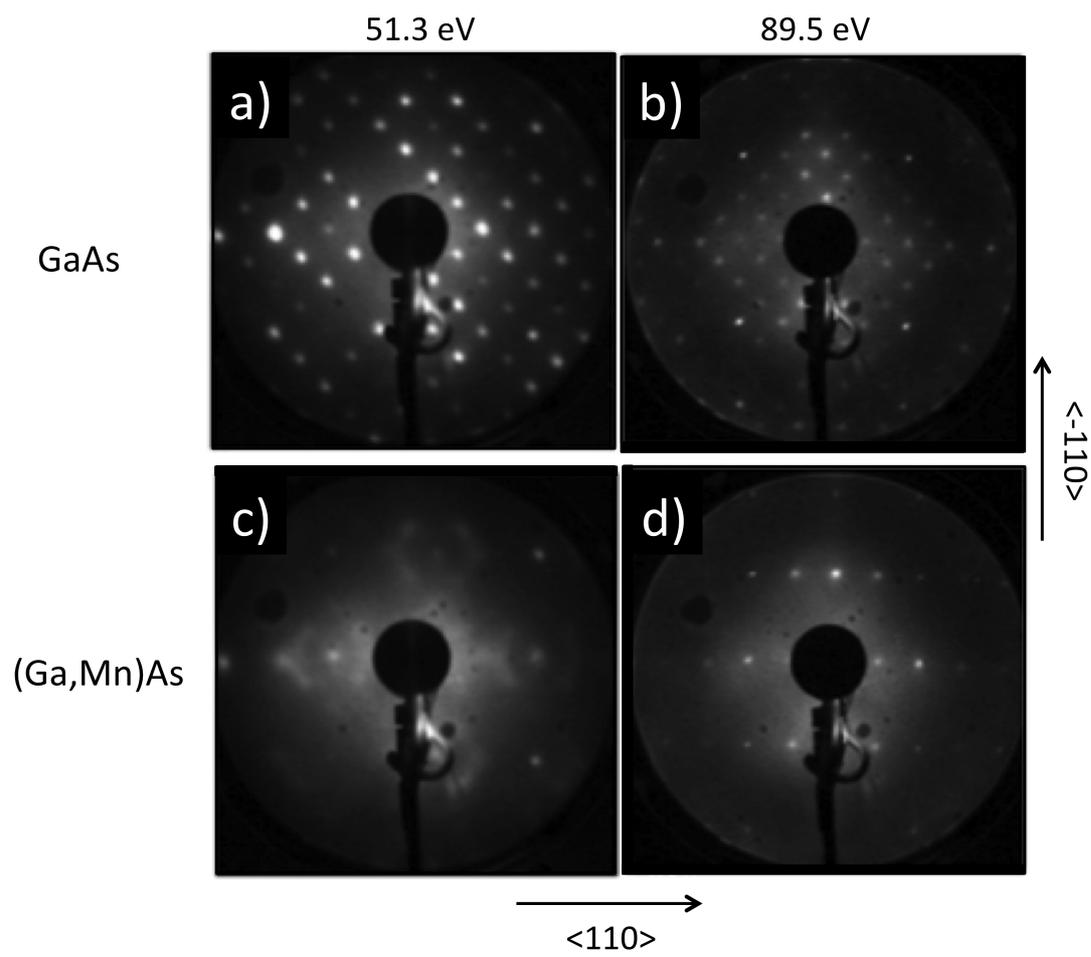

**Figure S1**

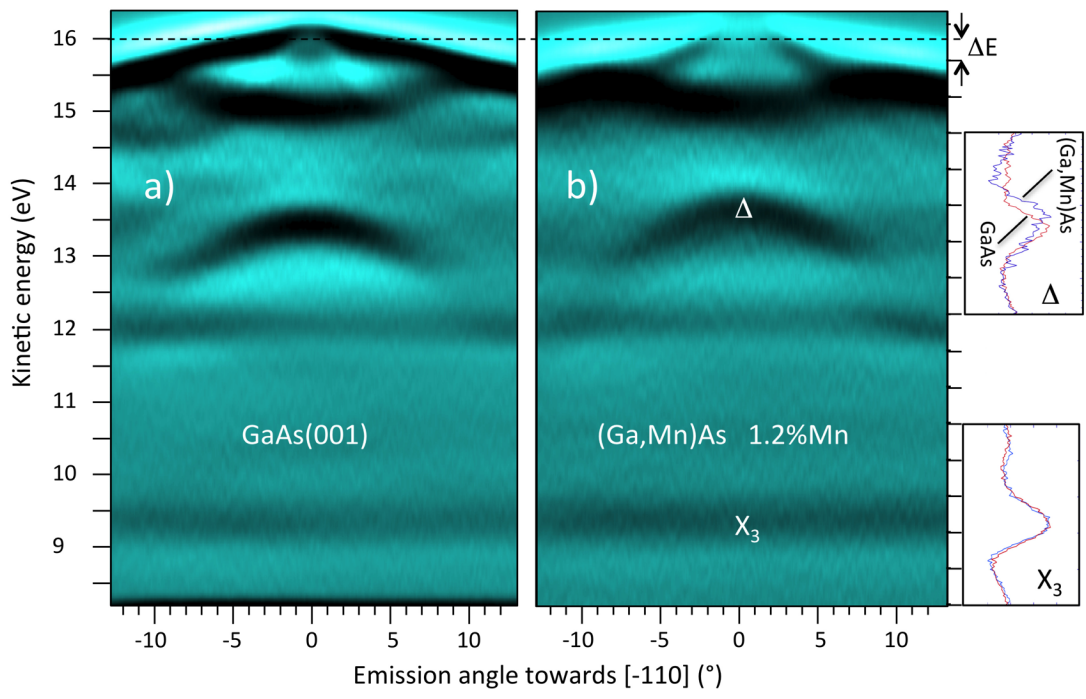

**Figure S2**

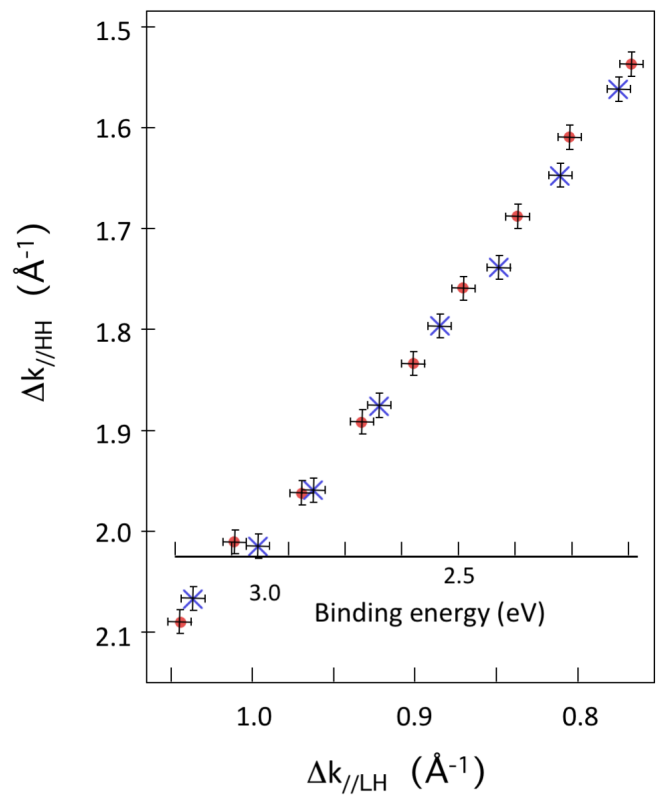

**Figure S3**